\documentclass[aip,aps,reprint,floatfix,twocolumn,,superscriptaddress]{revtex4}
\usepackage{graphicx}
\usepackage{amssymb}
\usepackage{amsmath}

\begin{document}

\title{Photoinduced valley polarized current of layered $\mathrm{MoS_2}$ by electric tuning}

\author{Yunjin Yu}
\affiliation{College of Physics Science and Technology, and Institute of Computational Condensed Matter Physics, Shenzhen University, Shenzhen 518060, P. R. China}
\author{Yanfeng Zhou}
\affiliation{International Center for Quantum Materials, School of Physics, Peking University, Beijing 100871, P. R. China}
\author{Langhui Wan}
\affiliation{College of Physics Science and Technology, and Institute of Computational Condensed Matter Physics, Shenzhen University, Shenzhen 518060, P. R. China}
\author{Bin Wang}
\affiliation{College of Physics Science and Technology, and Institute of Computational Condensed Matter Physics, Shenzhen University, Shenzhen 518060, P. R. China}
\author{Fuming Xu}
\affiliation{College of Physics Science and Technology, and Institute of Computational Condensed Matter Physics, Shenzhen University, Shenzhen 518060, P. R. China}
\author{Yadong Wei}
\email{ywei@szu.edu.cn}
\affiliation{College of Physics Science and Technology, and Institute of Computational Condensed Matter Physics, Shenzhen University, Shenzhen 518060, P. R. China}
\author{Jian Wang}
\email{jianwang@hku.hk}
\affiliation{Department of Physics, The University of Hong Kong, Hong Kong, P. R. China}

\date{\today}


\begin{abstract}
Photoinduced current of layered $\mathrm{MoS_2}$ based transistor is studied from first principles. Under the illumination
of  circular polarized light, valley polarized current is generated which can be tuned by the gate voltage.
For monolayer $\mathrm{MoS_2}$, the valley polarized spin up (down) electron current at $K$ ($K'$) points
is induced by the right (left) circular polarized light. The valley polarization is found to reach +1.0 (-1.0) for valley
current that carried such a $K$ ($K'$) index. For bilayer $\mathrm{MoS_2}$, the spin up (down)
current can be induced at both $K$ and $K'$ valleys by the right (left) circular light. In contrast to monolayer $\mathrm{MoS_2}$,
the photoinduced valley polarization shows asymmetric behavior upon reversal of the
gate voltage. Our results show that the valley polarization of photoinduced current
can be modulated by the circular polarized light and the gate voltage. All the results can be well
understood using a simple K$\cdot$P model.
\pacs{73.63.-b, 78.67.-n,  85.60.-q}
\end{abstract}

\maketitle
As one of the most promising two-dimensional materials, graphene has shown exceptional physical,
chemical, and optical properties.\cite{Singh,Andrei,Qiao,Bao} However, pristine graphene doesn't have gap between
valence band and conduction band which hampers its applications in semiconductor devices. Although
layered transition-metal dichalcogenide (TMDC) has the similar hexagonal structure like graphene,
it shows distinctly different properties from graphene.\cite{Radisavljevic,novoselov,CLee,Tkorn}
Firstly, TMDC has strong spin-orbit coupling which is originated from the $\it d$ orbitals of
the heavy metal atoms. This makes TMDC an exciting platform to explore spintronic
applications.\cite{ZYZhu,HMin,YYao} Secondly, monolayer TMDC, due to its inversion symmetry breaking,
displays distinct physical properties from its bulk counterpart. TMDC crossovers from an indirect
band gap semiconductor at bulk to a direct band gap semiconductor at monolayer.\cite{KFMak,ASplendiani}
Most importantly, monolayer TMDC has six valleys at the corners of its hexagonal Brillouin zone, which
can be classified into two inequivalent groups. Such valleys have large separations in momentum space
which makes the valley index robust against small deformation of its lattice
and low-energy scattering by long wavelength phonon. This means that the valley index can
be used as a potential information carrier. The valley properties of graphene has
been extensively studied theoretically.\cite{ARycerz,Dxiao,FZhang} Recently, there has been a growing
interest in the special spin and valley properties of layered TMDC both experimentally and
theoretically.\cite{AMJones,Dxiao2} Xiao {\it et al} \cite{Dxiao2} found that in monolayer TMDC,
inversion symmetry breaking and spin orbit coupling (SOC) lead to coupled spin and valley physics.
Monolayer TMDC has opposite spins at the two inequivalent $K$ points, making the optical
transition rules between valence band and conduction band both spin-dependent and valley-dependent.
Carriers with various combinations of valley and spin indices can be selectively excited
by optical fields with different circular polarizations.\cite{WYao,HYuan} Circularly polarized
luminescence has been observed in monolayer $\mathrm{MoS_2}$ and bilayer $\mathrm{MoS_2}$
under circularly optical pumping with different frequencies.\cite{Hzeng,Swu} This confirms the theoretical prediction that
the circular polarization originates from the contrasting selection rules for optical transitions in different valleys\cite{WYao}.

For the potential device application of layered TMDC, the key issue is to examine the performance of
nanoelectronic devices such as transistors. Indeed, the properties of electron-hole transport and photovoltaic
effect in gated $\mathrm{MoS_2}$ Schottky junction were studied.\cite{MFontana} The phototransistor
based on monolayer $\mathrm{MoS_2}$ exhibited good photoresponsivity and prompt photoswitching,
and the mechanism of photoresponse was analyzed in the ultrathin $\mathrm{MoS_2}$ field-effect
transistors by scanning photoinduced current microscopy.\cite{OLopez,RSsundaram,ZYin,CWu} So far, most
of the investigations concentrate on the I-V characteristics and less attention has been paid on the valley information.
Since one of the major challenges in valleytronics is the generation of valley polarized current,
it is important to study the valley polarized current through layered $\mathrm{MoS_2}$.
In this letter, we investigate the properties of valley polarized current of layered $\mathrm{MoS_2}$
phototransistor from first principles and analyze our results using K$\cdot$P model.

In this work, we calculate the photoinduced valley polarized current in monolayer and bilayer $\mathrm{MoS_2}$
phototransistors. As shown in Fig.1, the phototransistor consists of two semi-infinite sheets of
monolayer/bilayer $\mathrm{MoS_2}$ as leads and a central scattering region. A vertical electric field {\bf E}
along $z$ direction is produced by applying the gate voltage $V_g$ at the bottom gate in the central scattering region.
In the numerical calculation, the sizes of supercells used in our calculation are set to be
10.34 $a_B\times$ 47.77 $a_B \times$ 40.0 $a_B$ for monolayer $\mathrm{MoS_2}$ and 10.34 $a_B\times$ 47.77 $a_B\times$ 55.0 $a_B$ for bilayer $\mathrm{MoS_2}$. The band gaps we get here are $E^{mono}_{gap} = 1.7534$ eV for monolayer $\mathrm{MoS_2}$ and $E^{bi}_{gap} =1.7278$ eV for bilayer $\mathrm{MoS_2}$, which are similar to those of reference \onlinecite{Kadantsev} and reference \onlinecite{QihangLiu}, although they are under estimated comparing to the experimental results of reference \cite{KFMak}. We will focus on the transport along armchair direction when the circularly polarized light is shined on the central region. An external bias voltage $V_b = 0.3$ V is applied across the central region in order to collect valley polarized photo current. The energy of incident light is assumed to be equal to the direct energy gap $E^{mono}_{gap}$ for monolayer $\mathrm{MoS_2}$ and $E^{bi}_{gap}$ for bilayer $\mathrm{MoS_2}$, respectively.

To calculate the photoinduced current of such devices, we treat electron-photon coupling as a perturbation on the self
consistent Hamiltonian of electronic degrees of freedom $H_e$.\cite{RLake,LEHenrickson,JChen} To obtain the nonequilibrium
Hamiltonian $H_e$ of open structures, we employ the state-of-the-art first principles method based on the combination of density functional theory and
the Keldysh nonequilibrium Green's function formalism (NEGF-DFT).\cite{LKleinman}
The system nonequilibrium Hamiltonian $H_e$ is self-consistently determined through NEGF-DFT calculation
which includes spin-orbital coupling (SOI), external bias voltage and gate voltage.
Since the light consists of both electric and magnetic fields, the current density function theory (CDFT) may be more appropriate to describe the photocurrent.\cite{Vignale} For small field strength as is the case in this paper, DFT may be a good approximation.

Our calculation was preformed using the first principles package NanoDCal.\cite{JTaylor,Guohong}
Double-$\zeta$ basis set was used to expand the wave functions and the exchange-correlation potential
was treated at local spin density approximation (LSDA) level.\cite{JPPerdew,Kubler,Kubler2,Nordstrom} The mesh cut-off energy
was set to be 200 Ry and numerical tolerance of self-consistency was restricted to $10^{-4}$ eV. To consider the $k$-sampling, mesh $12 \times 1 \times 1$ in $k$-space is used.
After obtaining $H_e$, we treated the electron-photon interaction $H_{e-ph}$ by
the first order Born approximation. Here, $H_{e-ph}=\frac{e}{m}\mathbf{A}\cdot\mathbf{P}$, where $\mathbf{A}$ is the electromagnetic
vector potential and $\mathbf{P}$ the momentum of the electron. Detailed procedures of obtaining Green's function have been discussed in reference \onlinecite{Lzhang}.

\begin{figure}
\includegraphics[width=9cm,angle=0]{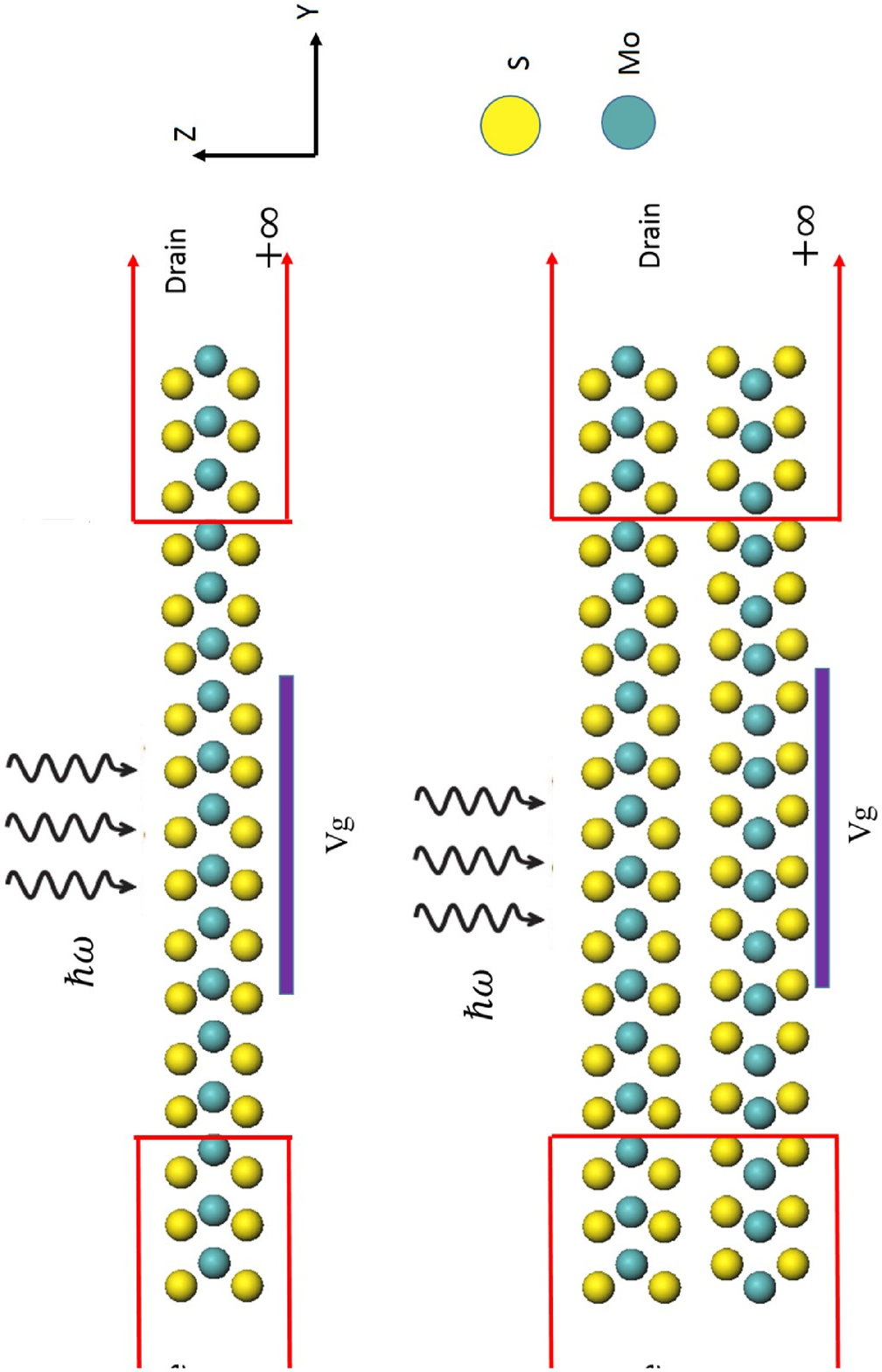}
\caption{Schematic plots of monolayer (upper) and bilayer (lower) $\mathrm{MoS_2}$ phototransistors.
Each central scattering region controlled by the gate voltage $V_{g}$ is sandwiched by source and
drain regions that extend to $y=\mp \infty$. The energy of shining light is expressed in terms of $\hbar\omega$.}
\label{fig1}
\end{figure}

The photoinduced valley and spin dependent current in the lead $\alpha$ can be written as\cite{Lzhang,HHaug}
\begin{equation}
I^{ph}_{\alpha,\tau,s}=\frac{e}{\hbar}\int\frac{dE}{2\pi}\sum_{{\bf k}\in\tau}T_{\alpha}^{ph}(E,{\bf k},s).
\end{equation}
Here,$\alpha=S/D$ stands for the lead of source/drain. $\tau=\pm1$ (corresponding to $K$/$K'$) is the valley index and
$\bf k$ around $K$ or $K'$ is calculated starting from the point of $K$ or $K'$, $s$ is the spin index.
$T_{\alpha}^{ph}$ is the effective transmission coefficient of lead $\alpha$, and its expression is
\begin{equation}
T_{\alpha}^{ph}(E,{\bf k},s)=\textrm{Tr}\{i\Gamma_{\alpha}(E,{\bf k})[(1-f_{\alpha})G_{ph}^{<}+f_{\alpha}G_{ph}^{>}]\}_{ss},
\end{equation}
where, $f_{\alpha}$ is the Fermi distribution function of lead $\alpha$, $\Gamma_{\alpha}$ is the linewidth
function which reflects the coupling between lead and central scattering region, $G^{<,>}_{ph}$ is the Green's function
including the contribution of voltage and photons.\cite{Lzhang}

\begin{figure}[tbp]
\includegraphics[width=8cm,angle=0]{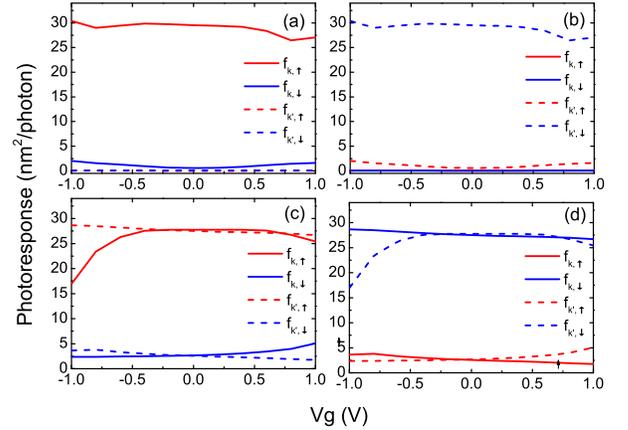}
\caption{The valley and spin components of photoresponse of monolayer (panel (a) and (b)) and bilayer
(panel (c) and (d)) $\mathrm{MoS_2}$ by {\it ab initio} method. The left two panels ((a) and (c)) correspond to the case of
right circular polarized light $\sigma^{+}$, the right two panels ((b) and (d)) correspond to the case of
left circular polarized light $\sigma^{-}$. }
\label{fig2}
\end{figure}

To describe the current response to the light, we examine the photoresponse, which is defined as
\begin{equation}
f_{\tau,s}=\frac{I^{ph}_{\tau,s}}{eF_{ph}},
\end{equation}
where $I^{ph}_{\tau,s}$ is the current with valley index $\tau$ and spin index $s$. $F_{ph}$ is the photon
flux defined as the number of photons per unit time per unit area.

Figure \ref{fig2} shows the valley and spin polarized photoresponse versus gate voltage under bias voltage 0.3 V.
The solid red line, solid blue line, dashed red line, and dashed blue line correspond to photoresponse
$f_{K,\uparrow}$, $f_{K,\downarrow}$, $f_{K',\uparrow}$, and $f_{K',\downarrow}$, respectively. From panel (a),
we see that the component $f_{K,\uparrow}$, is at least one order of magnitude larger than other three components,
indicating that the right circular polarized light $\sigma^{+}$ mainly excites the spin-up electrons at $K$ point from
the valence band to the conduction band in monolayer $\mathrm{MoS_2}$. Similarly we conclude from figure \ref{fig2}(b)
that the left circular polarized light $\sigma^{-}$ mainly excites the spin-down electrons at $K'$ point in monolayer
$\mathrm{MoS_2}$. This phenomenon is due to the symmetry breaking in monolayer $\mathrm{MoS_2}$ and can be explained
by optical selection rule.\cite{Dxiao2} The situation is different for bilayer $\mathrm{MoS_2}$ where the inversion
symmetry is restored. In this case, the spin up components of photonresponse ($f_{K,\uparrow}$ and $f_{K',\uparrow}$)
are much larger than the other two components when $\sigma^{+}$ light is shed (panel(c)), whereas the spin down
components ($f_{K,\downarrow}$ and $f_{K',\downarrow}$) dominate for $\sigma^{-}$ light (panel(d)).

\begin{figure}[tbp]
\includegraphics[width=9cm,angle=0]{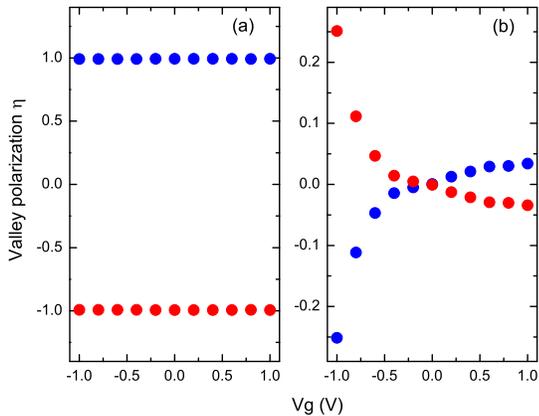}
\caption{The valley polarization versus gate voltage in monolayer (panel (a)) and
bilayer (panel(b)) $\mathrm{MoS_2}$ by {\it ab initio} method. The blue dot line corresponds to the case of $\sigma^{+}$ light and
the red dot line corresponds to the case of $\sigma^{-}$ light.}
\label{fig3}
\end{figure}

To characterize valley injection, we define valley polarization $\eta$ as
\begin{equation}
\eta=\frac{I^{ph}_{K}-I^{ph}_{K'}}{I^{ph}_{K}+I^{ph}_{K'}}
\end{equation}
where $I^{ph}_{K}=I^{ph}_{K\uparrow}+I^{ph}_{K\downarrow}$. From the panel (a) of figure \ref{fig3}, we see that
almost fully polarized valley current is generated using either $\sigma^{+}$ or $\sigma^{-}$ incident light.
In addition, the valley polarizations remain constant in the whole range of gate voltages from -1.0 V to 1.0 V,
suggesting that valley polarization of monolayer $\mathrm{MoS_2}$ is robust against the gate voltage. In contrast,
we find that the valley polarization is very sensitive to the gate voltage for bilayer $\mathrm{MoS_2}$. As shown in
panel (b) of figure 3, the valley polarization changes from -0.25 to 0.05 in bilayer $\mathrm{MoS_2}$ (blue dot line
in panel (b)) in the gate voltage window $[-1.0, 1.0]$ V when $\sigma^{+}$ light is shed. The valley polarization
profile $\eta_{\sigma}(V_g)$ satisfies $\eta_{+}(V_g)=-\eta_{-}(V_g)$. Hence for bilayer $\mathrm{MoS_2}$, the valley
polarization can be modulated by the circular polarized light as well as the gate voltage. We also find that the
modulation effect of negative gate voltage is more significant than that of positive one. To explain all of these
phenomena, we examine the following K$\cdot$P model.

Due to the inversion symmetry breaking and the presence of strong spin-orbit coupling in monolayer $\mathrm{MoS_2}$,
spin and valley degrees of freedom couple to each together, i.e, $K$ ($K'$) valley is occupied by the spin-up (down)
electrons at the top of the valence band. The electron interband transition from the top of spin-split valence band
to the bottom of the conduction band can be induced by the circular polarized light $\sigma^{\pm}$. If we define
the coupling strength with $\sigma^{\pm}$ optical fields as $\mathcal{P}_{\pm}({\bf k},s_z)$, we have the
following coupling intensity for transitions near $K/K'$ points\cite{Dxiao2}
\begin{equation}
|\mathcal{P}_{\pm}({\bf k},\tau,s_z)|^2=|P_0|^2(1\pm\tau\frac{\xi'}{\sqrt{\xi'^2+4a^2t^2k^2}}),
\label{coupling}
\end{equation}
where $|P_0|^2=\frac{m_{0}^{2}a^2t^2}{\hbar^2}$, $\xi'=\xi-\tau s_{z}\lambda$, $m_{0}$ is the
free electron mass, $a$ is the lattice constant, $t$ is the effective hopping integral,
$\xi$ is the energy gap, $\tau=\pm1$ is the valley index, 2$\lambda$ is the spin splitting
at the top of valence band caused by SOC,  and $s_{z}$ is for spin. To characterize the polarization
of coupling, we define
\begin{equation}
\eta^{c}_{\pm}=\frac{|\mathcal{P}_{\pm}(K)|^2-|\mathcal{P}_{\pm}(K')|^2}
{|\mathcal{P}_{\pm}(K)|^2+|\mathcal{P}_{\pm}(K')|^2}.
\label{cpeq}
\end{equation}
Here, $|\mathcal{P}_{\pm}(K/K')|^2=\sum_{s_z=\pm\frac{1}{2}} |\mathcal{P}_{\pm}(K/K',s_z)|^2$.
Since $k$ is very small near $K$ or $K'$ points, we have $\xi'\gg atk$.
This in turn gives $\eta^c_{\pm} = \pm 1$
from Eq. ({\ref{cpeq}}). The behavior of polarization of coupling
intensity is almost the same as that of valley current polarization obtained from {\it ab
initio} calculation shown in figure \ref{fig3}(a). This suggests that the polarization of valley current calculated
from ab initio method is intimately related to the polarization of coupling from the K$\cdot$P model in monolayer $\mathrm{MoS_2}$.

The situation is more complicated in bilayer $\mathrm{MoS_2}$. In this paper, we consider AB stacked bilayer $\mathrm{MoS_2}$
which maintains the inversion symmetry. To mimic the effect of the gate voltage in {\it ab initio} calculation, we introduce
an electric field along $z$ direction in our K$\cdot$P model Hamiltonian. The K$\cdot$P model Hamiltonian
for bilayer $\mathrm{MoS_2}$ with perpendicular external electric field is expressed as follows:\cite{GZRui}
\begin{equation}
  H({\bf k},\tau,s_z)=\begin{bmatrix}\begin{smallmatrix}
              \xi-\frac{\Delta U}{2}  & at(\tau k_{x}+ik_{y}) & 0      & 0    \\
    at(\tau k_{x}-ik_{y}) & -\tau s_{z}\lambda-\frac{\Delta U}{2}    & 0  & t_{\bot}  \\
    0   & 0   & \xi+\frac{\Delta U}{2}      & at(\tau k_{x}-ik_{y})\\
    0   & t_{\bot}  & at(\tau k_{x}+ik_{y})  & \tau s_{z}\lambda+\frac{\Delta U}{2}   \\
           \end{smallmatrix}\end{bmatrix} . \label{eq7}
\end{equation}
where $t_{\bot}$ is the intralayer hopping constant, $\Delta U=Ed$, $E$ is the magnitude of external electric field
and $d$ is the distance between two monolayers. The external electric field $E$ will induce an energy shift of
$\frac{-\Delta U}{2}$ at upper layer and an energy shift $\frac{\Delta U}{2}$ at lower layer in the bilayer $\mathrm{MoS_2}$.
Here all the parameters in Eq.(\ref{eq7}) are taken from the reference \onlinecite{GZRui} and the basis is
\{$|d^{u}_{z^2}>$,$\frac{1}{\sqrt{2}}(|d^{u}_{x^2-y^2}>-i\tau |d^{u}_{xy}>)$,$|d^{l}_{z^2}>$, $\frac{1}{\sqrt{2}}(|d^{l}_{x^2-y^2}>+i\tau |d^{l}_{xy}>)$\}.

By diagonalizing Eq. (\ref{eq7}), we obtain the eigenfunctions $|K\uparrow>$,
$|K\downarrow>$, $|K'\uparrow>$, and $|K'\downarrow>$ with ${\bf k}=0$ near the top valence band as follows,
\begin{eqnarray}
|K\uparrow>=\left(
  \begin{smallmatrix}
    0    \\
              \sin\alpha_{1}  \\
              0 \\
              \cos\alpha_{1}  \\
  \end{smallmatrix}
\right)\otimes|\uparrow>,
|K\downarrow>=\left(
  \begin{smallmatrix}
    0    \\
              \sin\alpha_{2}  \\
              0 \\
              \cos\alpha_{2}  \\
  \end{smallmatrix}
\right)\otimes|\downarrow>,\nonumber\\
|K'\uparrow>=\left(
  \begin{smallmatrix}
    0    \\
              \sin\alpha_{2}  \\
              0 \\
              \cos\alpha_{2}  \\
  \end{smallmatrix}
\right)\otimes|\uparrow>,
|K'\downarrow>=\left(
  \begin{smallmatrix}
    0    \\
              \sin\alpha_{1}  \\
              0 \\
              \cos\alpha_{1}  \\
  \end{smallmatrix}
\right)\otimes|\downarrow>
\end{eqnarray}
with
\begin{eqnarray}
\sin\alpha_{1}=\frac{t_{\perp}}{\sqrt{t_{\perp}^2+(E_{v1}+\lambda+\frac{\Delta U}{2})^2}},\nonumber\\
\cos\alpha_{1}=\frac{E_{v1}+\lambda+\frac{\Delta U}{2}}{\sqrt{t_{\perp}^2+(E_{v1}+\lambda+\frac{\Delta U}{2})^2}},\nonumber\\
\sin\alpha_{2}=\frac{E_{v2}+\lambda-\frac{\Delta U}{2}}{\sqrt{t_{\perp}^2+(E_{v2}+\lambda-\frac{\Delta U}{2})^2}},\nonumber\\
\cos\alpha_{2}=\frac{t_{\perp}}{\sqrt{t_{\perp}^2+(E_{v2}+\lambda-\frac{\Delta U}{2})^2}},\nonumber\\
\end{eqnarray}
and
\begin{eqnarray}
E_{v1}=\sqrt{t_{\perp}^2+(\lambda+\frac{\Delta U}{2})^2},\nonumber\\
E_{v2}=\sqrt{t_{\perp}^2+(\lambda-\frac{\Delta U}{2})^2}.
\end{eqnarray}

\begin{figure}[tbp]
\includegraphics[width=8cm,angle=0]{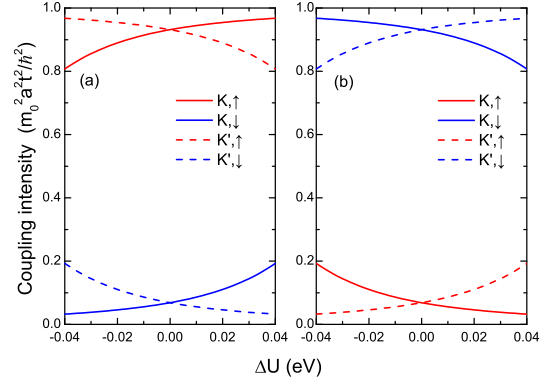}
\caption{The coupling intensity versus electron potential energy difference $\Delta U$ between the bottom
and top layers of bilayer $\mathrm{MoS_2}$. Panel (a) corresponds
to the case of $\sigma^{+}$ light and panel (b) corresponds to the case of $\sigma^{-}$ light. }
\label{fig4}
\end{figure}

Under the illumination of circular polarized light, the coupling intensities are found to be
\begin{eqnarray}
|\mathcal{P}_{+}(K,\uparrow)|^2=|P_0|^2\cos^2\alpha_{1},|\mathcal{P}_{+}(K,\downarrow)|^2=|P_0|^2\cos^2\alpha_{2},\nonumber\\
|\mathcal{P}_{-}(K,\uparrow)|^2=|P_0|^2\sin^2\alpha_{1},|\mathcal{P}_{-}(K,\downarrow)|^2=|P_0|^2\sin^2\alpha_{2},\nonumber\\
|\mathcal{P}_{+}(K',\uparrow)|^2=|P_0|^2\sin^2\alpha_{2},|\mathcal{P}_{+}(K',\downarrow)|^2=|P_0|^2\sin^2\alpha_{1},\nonumber\\
|\mathcal{P}_{-}(K',\uparrow)|^2=|P_0|^2\cos^2\alpha_{2},|\mathcal{P}_{-}(K',\downarrow)|^2=|P_0|^2\cos^2\alpha_{1}. \nonumber\\
\end{eqnarray}
Figure \ref{fig4} shows the coupling intensity versus the two layer potential energy difference $\Delta U$ in bilayer $\mathrm{MoS_2}$
by K$\cdot$P model. The results are similar to those of figure \ref{fig2} by the {\it ab initio} method.
Our {\it ab initio} calculation shows that $V_g = 1.0$ V gives rise a potential difference 0.04 V between the top and bottom layers
in bilayer $\mathrm{MoS_2}$. Hence we plot the coupling intensity versus $\Delta U$ from $\Delta U = -0.04$ V to $ 0.04$ V
in figure 4 in order to compare with figure 2, where $V_g$ is from -1.0 V to 1.0 V.

From Eq. (\ref{cpeq}), the valley coupling polarization for bilayer $\mathrm{MoS_2}$ is found to be

\begin{equation}
\eta^{c}_{\pm}=\pm\frac{1}{2}(\cos(2\alpha_{1})+\cos(2\alpha_{2}))
\label{coupol}
\end{equation}

\begin{figure}[tbp]
\includegraphics[width=9cm,angle=0]{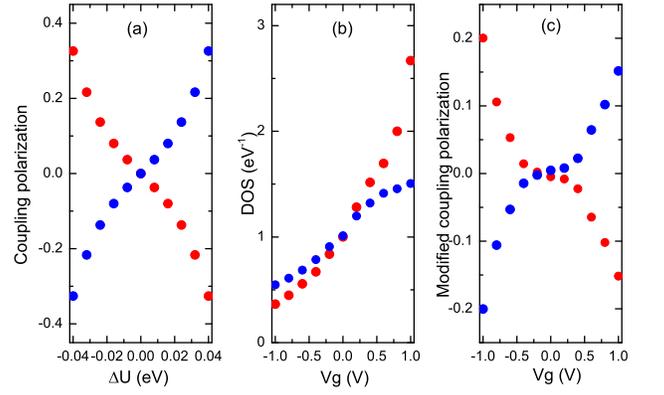}
\caption{ (a) The coupling polarization versus $\Delta U$ calculated from Eq. (\ref{coupol}) of K$\cdot$P model. The blue dotted
line is for the case of $\sigma^{+}$ light, the red dotted line for the case of $\sigma^{-}$ light.
(b) DOS of upper layer Mo atom (red dotted line) and lower layer Mo atom (blue dotted line) versus the gate voltage by {\it ab initio} method.
(c) The modified valley polarization versus the gate voltage (actually $\Delta U$) from Eq. (\ref{mvp}). The blue dotted line corresponds to
the case of $\sigma^{+}$ light, the red dotted line to the case of $\sigma^{-}$ light. }
\label{fig5}
\end{figure}

In figure \ref{fig5}(a), we plot the valley coupling polarization versus the gate voltage using the formula above.
Similar to figure \ref{fig2}(b), we have $\eta^c_+(\Delta U) =-\eta^c_-(\Delta U)$.
However, we also have a relation $\eta^c_{\pm}(\Delta U) =-\eta^c_{\pm}(-\Delta U)$, i.e., $\eta^c_{\pm}$ are
odd functions of $\Delta U$ which are different from our first principles results.
In fact, from Eq. (\ref{coupol}) one can easily get $\eta^c_\pm \to \pm \frac{t^2_\perp}{(\lambda^2 + t^2_{\perp})^{3/2}} \Delta U$
when $\Delta U \to 0$. To understand this difference, we examine the contribution from DOS of the valence band which
is not considered in K$\cdot$P model. Since the photoinduced current originates from the transition between different
valence bands to the same conduction band we will neglect the influence of DOS of the conduction band.
DOS near the top of valence band will be affected by the external electric field.
Our {\it ab initio} results show that DOS of valence band are mainly contributed by Mo atoms of bilayer $\mathrm{MoS_2}$.
From analytic calculation, we find that $|\mathcal{P}_{+}(K,s_z)|^2$ ($|\mathcal{P}_{-}(K,s_z)|^2$) is related mainly to
DOS from lower (upper) layer of bilayer $\mathrm{MoS_2}$ and $|\mathcal{P}_{+}(K',s_z)|^2$ ($|\mathcal{P}_{-}(K',s_z)|^2$)
is related mainly to DOS from upper (lower) layer of bilayer $\mathrm{MoS_2}$. In figure \ref{fig5}(b), we plot the DOS of
upper layer Mo atom (red dotted line) and lower layer Mo atom (blue dotted line) versus gate voltage obtained by {\it ab}
initio method. From this figure, we see that the influence of external electric field on DOS of Mo atom of upper layer is
different from that of lower layer. In another word, a better definition of valley coupling polarizations should include
the effect of DOS as follows,
\begin{equation}
\eta^{c}_{\pm}=\pm\frac{D_{l}\cos^2\alpha_{1}-D_{u}\sin^2\alpha_{1}+D_{l}\cos^2\alpha_{2}-D_{u}\sin^2\alpha_{2}}
{D_{l}\cos^2\alpha_{1}+D_{u}\sin^2\alpha_{1}+D_{l}\cos^2\alpha_{2}+D_{u}\sin^2\alpha_{2}}.
\label{mvp}
\end{equation}
where $D_{l}$ and $D_{u}$ are DOS of lower and upper Mo atoms, respectively. In panel (c) of
figure \ref{fig5}, we plot the modified coupling polarization as a function of $\Delta U$ using Eq. (\ref{mvp}) with DOS
taken from the panel (b) of figure \ref{fig5}. In order to compare with figure \ref{fig3}, we changed the abscissa of
panel (c) from $\Delta U$ from -0.04 V to 0.04 V to $V_g$ from -1.0 V to 1.0 V according to our {\it ab initio} results. We see that the
modified coupling polarization is no longer an odd function of the gate voltage, and the behaviors of $\eta^c_{\pm}(V_g)$ for $V_g=[-1,0.3]$ V
are similar to the results in figure 3(b). In another word, by analyzing the polarization of coupling, one can get the information of the valley
polarization. For the monolayer $\mathrm{MoS_2}$, one can directly using the K$\cdot$P model, but for the bilayer $\mathrm{MoS_2}$, one has to
consider the influence of the DOS of the energy bands.

In summary, we have investigated the photoinduced current of layered $\mathrm{MoS_2}$
as a function of external electric field. The results show that the valley
polarization of photoinduced current of monolayer $\mathrm{MoS_2}$ is independent of
external electric field perpendicular to the surface of the layered $\mathrm{MoS_2}$
which can be induced by the gate voltage, but the valley polarization of photoinduced current of
bilayer $\mathrm{MoS_2}$ is very sensitive to the external electric field that breaks
the inversion symmetry. Moreover, the valley polarization can be tuned by changing the polarity
of circular polarized light. The modulation of valley polarization of layered $\mathrm{MoS_2}$
transistor by gate voltages and polarities of circular polarized light provide extra knot in future
application of valleytronic devices.

\bigskip
This work was supported by National Natural Science Foundation of China (No.11374246, No.11574217, and No.11504240
) and Shenzhen
Natural Science Foundation (JCYJ20150324140036832).

\end{document}